\documentclass[prl,showpacs,twocolumn,aps,superscriptaddress,a4paper]{revtex4-1}

\usepackage{pst-node}
\usepackage{amsmath}
\usepackage{graphicx}
\usepackage{latexsym}
\usepackage{amsfonts}
\usepackage{amssymb}
\usepackage{graphicx}
\usepackage{rotating}
\usepackage{bbm}
\usepackage{cancel}

\DeclareGraphicsExtensions{.pdf,.gif,.jpg,.png}

\newcommand{\Tr}{ {\rm Tr} \, }
\newcommand{\be}{\begin{equation}}
\newcommand{\ee}{\end{equation}}
\newcommand{\bea}{\begin{eqnarray}} 
\newcommand{\eea}{\end{eqnarray}}

\begin{document}  

\def\gC{\mbox{\boldmath $C$}}
\def\gZ{\mbox{\boldmath $Z$}}
\def\gR{\mbox{\boldmath $R$}}
\def\gN{\mbox{\boldmath $N$}}
\def\gG{\mbox{\boldmath $G$}}
\def\green{\mbox{\boldmath ${\cal G}$}}
\def\grn{\mbox{${\cal G}$}}
\def\gH{\mbox{\boldmath $H$}}
\def\bA{{\bf A}}
\def\bJ{{\bf J}}
\def\bG{{\bf G}}
\def\bF{{\bf F}}
\def\bH{\mbox{\boldmath $H$}}
\def\bQ{\mbox{\boldmath $Q$}}
\def\bgS{\mbox{\boldmath $\Sigma$}}
\def\bT{\mbox{\boldmath $T$}}
\def\bU{\mbox{\boldmath $U$}}
\def\bV{\mbox{\boldmath $V$}}
\def\bgG{\mbox{\boldmath $\Gamma$}}
\def\bgL{\mbox{\boldmath $\Lambda$}}
\def\ubG{\underline{{\bf G}}}
\def\ubH{\underline{{\bf H}}}
\def\ubQ{\underline{{\bf Q}}}
\def\ubS{\underline{{\bf S}}}
\def\ubg{\underline{{\bf g}}}
\def\ubq{\underline{{\bf q}}}
\def\ubp{\underline{{\bf p}}}
\def\ubgS{\underline{{\bf \Sigma}}}
\def\bge{\mbox{\boldmath $\epsilon$}}
\def\bgD{{\bf \Delta}}

\def\bDelta{\mbox{\boldmath $\Delta$}}
\def\bcalE{\mbox{\boldmath ${\cal E}$}}
\def\bcalF{\mbox{\boldmath ${\cal F}$}}
\def\bcalG{\mbox{\boldmath $G$}}
\def\ubcalG{\mbox{\underline{\boldmath $G$}}}
\def\callG{\mathcal{G}}
\def\callA{\mathcal{A}}
\def\callT{\mathcal{T}}
\def\ubcalA{\mbox{\underline{\boldmath $A$}}}
\def\ubcalB{\mbox{\underline{\boldmath $B$}}}
\def\ubcalC{\mbox{\underline{\boldmath $C$}}}
\def\ubcalg{\mbox{\underline{\boldmath $g$}}}
\def\ubcalH{\mbox{\underline{\boldmath $H$}}}
\def\bcalK{\mbox{\boldmath $K$}}
\def\ubcalK{\mbox{\underline{\boldmath $K$}}}
\def\bcalV{\mbox{\boldmath ${\cal V}$}}
\def\ubcalV{\mbox{\underline{\boldmath $V$}}}
\def\bcalU{\mbox{\boldmath ${\cal U}$}}
\def\ubcalz{\mbox{\underline{\boldmath $z$}}}
\def\bcalQ{\mbox{\boldmath ${\cal Q}$}}
\def\ubcalQ{\mbox{\underline{\boldmath $Q$}}}
\def\ubcalP{\mbox{\underline{\boldmath $P$}}}
\def\bSS{\mbox{\boldmath $S$}}
\def\ubcalS{\mbox{\underline{\boldmath $S$}}}
\def\bff{\mbox{\boldmath $f$}}
\def\bg{\mbox{\boldmath $g$}}
\def\bh{\mbox{\boldmath $h$}}
\def\bk{\mbox{\boldmath $k$}}
\def\bq{\mbox{\boldmath $q$}}
\def\bp{\mbox{\boldmath $p$}}
\def\br{\mbox{\boldmath $r$}}
\def\bt{\mbox{\boldmath $t$}}
\def\ubh{\mbox{\underline{\boldmath $h$}}}
\def\ubt{\mbox{\underline{\boldmath $t$}}}
\def\ubk{\mbox{\underline{\boldmath $k$}}}
\def\ua{\uparrow}
\def\da{\downarrow}
\def\a{\alpha}
\def\b{\beta}
\def\g{\gamma}
\def\G{\Gamma}
\def\d{\delta}
\def\D{\Delta}
\def\e{\epsilon}
\def\ve{\varepsilon}
\def\z{\zeta}
\def\h{\eta}
\def\th{\theta}
\def\vth{\vartheta}
\def\k{\kappa}
\def\l{\lambda}
\def\L{\Lambda}
\def\m{\mu}
\def\n{\nu}
\def\x{\xi}
\def\X{\Xi}
\def\p{\pi}
\def\P{\Pi}
\def\r{\rho}
\def\bgr{\mbox{\boldmath $\rho$}}
\def\s{\sigma}
\def\us{\mbox{\underline{\boldmath $\sigma$}}}
\def\ubgm{\mbox{\underline{\boldmath $\mu$}}}
\def\S{\Sigma}
\def\ubcgS{\mbox{\underline{\boldmath $\Sigma$}}}
\def\t{\tau}
\def\f{\phi}
\def\vf{\varphi}
\def\F{\Phi}
\def\c{\chi}
\def\k{\kappa}
\def\w{\omega}
\def\W{\Omega}
\def\Q{\Psi}
\def\q{\psi}
\def\de{\partial}
\def\inf{\infty}
\def\ra{\rightarrow}
\def\bra{\langle}
\def\ket{\rangle}
\def\bbra{\langle\langle}
\def\kket{\rangle\rangle}
\def\grad{\mbox{\boldmath $\nabla$}}
\def\no{\bf 1}
\def\ze{\bf 0}
\def\uno{\underline{\bf 1}}
\def\zero{\underline{\bf 0}}

\def\dr{{\rm d}}
\def\bj{{\bf j}}
\def\br{{\bf r}}
\def\bz{\bar{z}}
\def\bart{\bar{t}}

\title{Steady-State Density Functional Theory for Finite Bias Conductances}

\author{G. Stefanucci}
\affiliation{Dipartimento di Fisica, Universit\`{a} di Roma Tor 
Vergata and European Theoretical Spectroscopy Facility (ETSF),
Via della Ricerca Scientifica 1, 00133 Rome, Italy}
\affiliation{INFN, Laboratori Nazionali di Frascati, Via E. Fermi 40, 00044 Frascati,
Italy}

\author{S. Kurth}
\affiliation{Nano-Bio Spectroscopy Group and European Theoretical Spectroscopy Facility (ETSF),
Dpto. de F\'{i}sica de Materiales,
Universidad del Pa\'{i}s Vasco UPV/EHU, Av. Tolosa 72, 
E-20018 San Sebasti\'{a}n, Spain}
\affiliation{IKERBASQUE, Basque Foundation for Science, Maria Diaz de Haro 3, 
E-48013 Bilbao, Spain}

\begin{abstract}
In the framework of density functional theory a formalism to describe 
electronic transport in the steady state is proposed which uses the 
density on the junction and the {\em steady current} as basic variables.  
We prove that, in a finite window around zero bias, there is a one-to-one map 
between the basic variables and both local potential on as well as bias 
across the junction. The resulting Kohn-Sham system features two 
exchange-correlation (xc) potentials, a local xc potential and an xc 
contribution to the bias. For weakly coupled junctions the 
xc potentials exhibit steps in the density-current plane
which  are shown to be crucial to describe the Coulomb blockade 
diamonds. At small currents these 
steps emerge as the
equilibrium xc discontinuity bifurcates. The formalism is applied to 
a model benzene junction, finding perfect agreement with 
the orthodox theory of Coulomb blockade.

\end{abstract}


\maketitle

Engineering electrical transport through molecular junctions is a 
mandatory passage toward the miniaturization and speeding up of
device components \cite{me-book1,me-book2}. As a systematic experimental 
characterization of every synthesizable molecule is impractical, 
reliable theoretical methods are of utmost importance to 
progress. Density Functional 
Theory (DFT) has emerged as the method of choice due to 
the chemical complexity of the junctions
\cite{tgw.2001,bmots.2002,PdC.2004,rgblfs.2006,awe.2007}. Nevertheless, to date there 
exists still no DFT scheme to deal with the ubiquitous Coulomb 
Blockade (CB) phenomenon. CB stems from the interplay between quantum 
confinement and Coulomb repulsion, and it leaves clear fingerprints in 
the measured conductances. 

As recognized by several authors \cite{tfsb.2005,kbe.2006,kskvg.2010,ks.2013,lb.2015},  the 
discontinuity of the exchange-correlation (xc) potential plays a 
pivotal role in blocking 
the electron density at integer values. Although at low temperatures 
and for single-channel junctions this key xc feature yields also accurate 
DFT conductances \cite{sf.2008,mera-1,mera-2,sk.2011,blbs.2012,tse.2012}, 
we showed that at finite temperature 
and/or bias the {\em exact} DFT conductance does not exhibit any CB 
signature \cite{ks.2013}. In fact, static DFT misses the dynamical 
xc bias corrections~\cite{sa-1.2004,sa-2.2004,ewk.2006,szvv.2005} 
predicted by Time-Dependent (TD) DFT~\cite{rg.1984}, the proper framework 
within which to formulate a theory of quantum transport 
\cite{sa-1.2004,sa-2.2004}. Only 
recently a dynamical xc correction has been proposed \cite{ks.2013}, but
its applicability is limited to the CB regime at zero bias. 

In this Letter we put forward a steady-state DFT, henceforth named 
i-DFT, whose xc potentials 
(gate and bias) are functionals of the molecular density and the 
{\em steady-state} current. The i-DFT framework is suited to 
study the finite-bias and finite-temperature conductance as function of external gate 
and applied bias, and it generalizes standard DFT 
in equilibrium. Unlike Current DFT~\cite{vr.1987} (CDFT), i-DFT applies out of 
equilibrium and unlike TD(C)DFT~\cite{gd.1988,vk.1996} the i-DFT functional  
has no memory. The empirical Landauer+DFT approach to 
transport~\cite{tgw.2001,bmots.2002,PdC.2004,rgblfs.2006,awe.2007} 
is recovered as an approximation to i-DFT with the xc bias set to 
zero.
As we shall see this approximation is too severe in the CB regime.

Through a reverse-engineering procedure we show that 
the well-known discontinuities of the DFT xc potential at integer 
particle number 
$N$ {\em bifurcate} as the current $I$ starts flowing. The xc gate 
and xc bias 
exhibit an intricate and, at first sight, inexplicable pattern of intersecting 
steps in the $N$-$I$ plane. We recognize, however, that every 
intersection occurs at the plateau values of $N$ and $I$ in a CB diamond. 
This ``duality'' between intersections and  CB plateaus in current and 
particle number is an exact property of the xc potentials of 
nonequilibrium open systems and the 
fundamental ingredient to extend DFT transport calculations to finite 
bias.

\begin{figure}[tbp]
\includegraphics[width=0.48\textwidth]{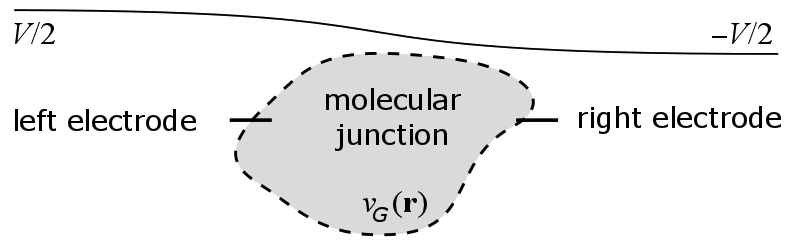}
\caption{Schematic illustration of a molecular 
junction attached to a left and right electrode. The system 
is subject to a bias voltage $V$ (solid line) and a gate potential 
$v_{G}(\br)$. The grey area is the molecular region defined in the main 
text.}
\label{junction}
\end{figure}

{\bf i-DFT} We consider a current-carrying molecular junction attached to a left ($L$) and right 
($R$) electrode in the {\em steady state}, see Fig.~\ref{junction}. 
In addition to the nuclear potential $v_{n}(\br)$
the electrons are subject to an external bias $v_{B}(\br)$ generated by 
a battery and to an external gate $v_{G}(\br)$ that vanishes 
deep inside the electrodes.
The classical potential $v_{n}(\br)+v_{B}(\br)+v_{H}(\br)$, 
$v_{H}$ being the Hartree potential,
in the far $L$ ($R$) region differs by a uniform 
shift $V/2$ ($-V/2$) from its equilibrium value, $V$ being the applied 
bias.\cite{asymbias} Since a change 
of the external gate 
does not affect the value of $V$ we can unambiguously calculate 
the steady-state current and density by specifying $v_{n}(\br)$, $v_{G}(\br)$ and 
$V$.~\cite{sa-1.2004,sa-2.2004} 

Let us select an arbitrary {\em finite} region of space 
$\mathcal{R}$ around the molecule, 
henceforth named {\em molecular region}, and write the nuclear/gate 
potential as
$v_{n/G}(\br)=v^{\rm in}_{n/G}(\br)+v^{\rm out}_{n/G}(\br)$ where 
$v_{n/G}^{\rm in}(\br)=v_{n/G}(\br)$ for $\br\in \mathcal{R}$ and zero otherwise. 
Similarly, we write the electronic density as $n(\br)=
n^{\rm in}(\br)+n^{\rm out}(\br)$ where $n^{\rm in}(\br)=n(\br)$ for 
$\br\in \mathcal{R}$ and zero otherwise.
We advance that for practical applications a convenient choice of 
the molecular region is the one for which 
in the electrodes (outside $\mathcal{R}$) 
we have $v_{G}\simeq 0$  and 
$v_{n}+v_{H}$  weakly dependent on the chemical structure of the junction. 
In this case  $n^{\rm out}(\br)$ is 
not affected by a change in $v_{G}$, a condition used in all DFT-based quantum 
transport calculations.
However, the formal results contained in this
section are independent of the choice of $\mathcal{R}$.

The foundation of i-DFT rests on the one-to-one 
correspondence between the two pairs 
$(v^{\rm in}(\br), V)$ and $(n^{\rm in}(\br), I)$. 
The first pair consists of the molecular potential 
$v^{\rm in}(\br)\equiv v_{n}^{\rm in}(\br)+v_{G}^{\rm in}(\br)$ and 
the bias $V$ whereas the 
second pair consists of the molecular density $n^{\rm in}(\br)$
and the steady current $I$.

{\em Theorem}.-- For any {\em finite} temperature and  at {\em fixed} 
outer potential
$v^{\rm out}(\br)\equiv v_{G}^{\rm out}(\br)+v_{n}^{\rm out}(\br)$ 
the map $(v^{\rm in}(\br),V) \rightarrow (n^{\rm in}(\br),I)$ is invertible 
in a finite bias window around $V=0$.

{\em Proof}.-- To prove the theorem we show that the Jacobian 
\be
J_{V=0}={\rm Det}
\left[
\begin{array}{cc}
    \d n^{\rm in}(\br) /\d v^{\rm in}(\br') &  \de n^{\rm in}(\br)/\de V \\ 
   \d I/\d v^{\rm in}(\br') &  \de I/\de V 
\end{array}
\right]_{V=0}.
\label{JV}
\ee
is nonvanishing (we are of course working under the physically 
reasonable assumption that $n^{\rm in}(\br)$ and $I$ are 
continuously differentiable, which also implies that $J_{V}$ is continuous 
in $V=0$).  
The block 
$\chi^{\rm in}(\br,\br')\equiv \d n^{\rm in}(\br) /\d v^{\rm in}(\br')|_{V=0}$ is the 
static {\em equilibrium} density response function of the 
contacted system with 
$\br,\br'$ in the molecular region,  whereas $G\equiv\de I/\de 
V|_{V=0}$ is the zero-bias conductance. 
We start by showing that $\chi^{\rm in}$ is invertible for any 
molecular region $\mathcal{R}$. The equilibrium response function 
$\chi^{\rm in}$ can be calculated using leads of finite 
length $L$ and then taking the limit $L\to\infty$.
Let $\{|\Q_{i}\ket\}$ be a complete set of many-body eigenstates of the 
equilibrium {\em contacted} system with 
energy $E_{i}$ and number of particles $N_{i}$. At temperature $1/\b$ 
and chemical potential $\m$ the  Lehmann representation of $\chi^{\rm in}$ reads~\cite{svl-book}
\be
\chi^{\rm in}(\br,\br')=\frac{1}{Z}\sum_{ij}
\frac{f_{ij}(\br)f_{ij}(\br')}{\W_{ij}^{2}+\eta^{2}}\W_{ij}
\left(e^{-\b E_{i}}-e^{-\b E_{j}}\right)e^{\b\m N_{i}}
\label{lehmannchi}
\ee
with $Z$ the partition function, $\W_{ij}=E_{i}-E_{j}$ the energy 
difference, $\eta$ a positive infinitesimal 
to set to zero after the limit $L\to\infty$, 
and $f_{ij}(\br)=\bra\Q_{i}|\hat{n}(\br)|\Q_{j}\ket-\d_{ij}n(\br)$ the excitation 
amplitudes. 
We define $T_{ij}\equiv\int_{\mathcal{R}} d\br 
f_{ij}(\br)t(\br)$ where $t(\br)$ is a test function. Notice that 
$T_{ij}$ has a well defined limit for $L\to\infty$ since the 
integral is over the finite domain $\mathcal{R}$. Proving the invertibility 
of $\chi^{\rm in}$ is equivalent to proving that
\bea
\int_{\mathcal{R}} d\br d\br' t(\br)\chi^{\rm in}(\br,\br')t(\br')=
\frac{1}{Z}\sum_{ij}
\frac{|T_{ij}|^{2}}{\W_{ij}^{2}+\eta^{2}}\W_{ij}
\nonumber \\
\times
\left(e^{-\b E_{i}}-e^{-\b E_{j}}\right)e^{\b\m N_{i}}\neq 0
\label{poschi}
\eea
for any test function $t(\br)$. To this end we observe that  
for every $E_{i}\lessgtr E_{j}$ we have $\W_{ij}\lessgtr 0$ and 
$(e^{-\b E_{i}}-e^{-\b E_{j}})\gtrless 0$. 
Hence the left hand side of Eq.~(\ref{poschi}) is non-positive 
and can be zero only provided that $T_{ij}=0$ for every $E_{i}\neq E_{j}$. 
However, this latter circumstance implies that the Hamiltonian 
and the operator $\hat{T}\equiv\int_{\mathcal{R}} 
d\br\,\hat{n}(\br)t(\br)$ can be diagonalized simultaneously, an 
absurdum for any test function $t(\br)$. Therefore Eq.~(\ref{poschi}) 
holds true and $\chi^{\rm in}$ is invertible.
Similarly, from the  Lehmann representation of the zero-bias 
conductance~\cite{bsw.2006} 
\be
G=-\frac{1}{Z}\sum_{ij}
\frac{2\eta |I_{ij}|^{2}}{(\W_{ij}^{2}+\eta^{2})^{2}}\W_{ij}
\left(e^{-\b E_{i}}-e^{-\b E_{j}}\right)e^{\b\m N_{i}}
\label{lehmanng}
\ee
one finds the intuitive result $G>0$. 
In Eq.~(\ref{lehmanng}) 
$I_{ij}\equiv \bra\Q_{i}|\hat{I}|\Q_{j}\ket$ with $\hat{I}$ the 
longitudinal current operator.
The crucial observation to end the proof is that 
at zero bias a variation of $v^{\rm in}$ does 
not induce a steady current, hence $\d I/\d v^{\rm in}(\br')|_{V=0} =0$. 
We conclude that  $J_{0}={\rm Det}[\chi^{\rm in}] G<0$ for all 
$v^{\rm in}$. 
Since $J_{V}$ is a continuous function of $V$ around $V=0$, 
there exists a finite interval (depending on $v^{\rm in}$) around $V=0$ for 
which $J_{V}<0$. In this domain the map $(v^{\rm in}(\br),V) \rightarrow 
(n^{\rm in}(\br),I)$ is invertible. 

An interesting consequence of the i-DFT theorem is that at zero bias 
(and hence at zero current) it 
generalizes standard equilibrium DFT at finite temperatures 
\cite{m.1965} to portions of an interacting system.
In fact, 
the i-DFT theorem implies that two potentials $v$ 
and $v'$ differing only in a region $\mathcal{R}$ 
generate two  equilibrium densities $n(\br)$ and $n'(\br)$ which 
are certainly different in $\mathcal{R}$ (see also 
Ref.~\citenum{rm.1981}). It is worth 
observing that the zero-bias i-DFT does not suffer from the DFT problem 
of infinite systems~\cite{ggg.1995} since the map  involves only the 
density and the potential in  $\mathcal{R}$. Furthermore,
since we are not interested in the density outside $\mathcal{R}$, 
no assumption on the analyticity of the density in position space is 
needed \cite{zwymc.2007}.
Interestingly, the zero-bias i-DFT for infinite 
systems can easily be generalized to the time domain too. Indeed, the boundary term in the 
Runge-Gross proof~\cite{rg.1984} vanishes identically since $v-v'$ is, by definition, zero 
outside $\mathcal{R}$.

Henceforth we omit the superscript ``in'' in the 
molecular density and potential; thus $n(\br)$ and $v(\br)$ are always 
calculated in $\br\in \mathcal{R}$. 
Let $(n,I)$ be the molecular density and current induced by the 
potentials $(v,V)$ in an {\em interacting} junction. 
Under the usual assumption of non-interacting $v$-representability, 
the i-DFT theorem guarantees that a pair of potentials $(v_{s},V_{s})$ which 
reproduces the same $(n,I)$ in a {\em noninteracting} 
junction is unique. Following the Kohn-Sham (KS)
procedure we define the xc bias and Hartree-xc (Hxc) gate as
\begin{subequations}
\bea
V_{\rm xc}[n,I]&\equiv&V_{s}[n,I]-V[n,I],
\\
v_{\rm Hxc}[n,I](\br)&\equiv&v_{s}[n,I](\br)-v[n,I](\br).
\eea
\label{xcpot}
\end{subequations}

\noindent
The nomenclature Hxc gate instead of Hxc 
molecular potential is used for brevity to indicate that $v_{\rm Hxc}$ is nonzero only in the molecular 
region.
The self-consistent KS equations then read (hereafter $\int\equiv \int \frac{d\w}{2\p}$)
\begin{subequations}
\bea
n(\br)&=&2\sum_{\a=L,R}\int f\big(\w+s_{\a}\frac{V+V_{\rm xc}}{2}\big) 
A_{\a}(\br,\w),
\label{ksn}
\\
I&=&2\sum_{\a=L,R}\int f\big(\w+s_{\a}\frac{V+V_{\rm 
xc}}{2}\big)s_{\a}T(\w).\quad\;\;
\label{ksi}
\eea
\label{ks}
\end{subequations}

\noindent
In Eqs.~(\ref{ks}), $f(\w)=1/(e^{\b(\w-\m)}+1)$ is the Fermi function 
whereas
$s_{R/L}=\pm$. We also defined the partial spectral function $A_{\a}(\br,\w)\equiv 
\bra\br|\callG(\w)\G_{\a}(\w)\callG^{\dag}(\w)|\br\ket$, with 
$\callG$ and $\G_{\a}$ the KS Green's function and broadening 
matrices~\cite{svl-book}, and the transmission function 
$T=\Tr[\callG(\w)\G_{L}(\w)\callG^{\dag}(\w)\G_{R}(\w)]$. 
In Eqs.~(\ref{ks}) one should note the presence of the 
KS bias $V_s=V+V_{\rm xc}$ instead of the bare bias $V$; this comes 
from the mapping $(n,I)\leftrightarrow (v_{s},V_{s})$.
Although the KS Eqs.~(\ref{ks}) for $n$ and $I$ are formally identical to 
the TDDFT expressions of Ref.~\citenum{sa-1.2004,sa-2.2004}, we stress that 
the i-DFT xc potentials depend on the {\em steady-state} molecular density 
and current while the TDDFT xc 
potential depend on the full history of the molecular {\em and} lead 
densities. The augmented local 
character of the i-DFT xc potentials is in agreement with similar 
findings in TDCDFT \cite{v.1995,nptvc.2007}. 

The simplifications brought about by i-DFT are especially 
evident when considering the zero-bias conductance. From Eq. 
(\ref{ksi}) we have
\be
G=G_{s}
\left[1+\frac{\de V_{\rm xc}}{\de I}G+
\int \!d\br\, \frac{\d V_{\rm xc}}{\d 
n(\br)}\frac{\de n(\br)}{\de V}\right]_{V=0},
\ee
with $G_{s}\equiv-2\int f'(\w)T(\w)$ the zero-bias KS conductance.
Since $I=0$ is the 
solution of the KS equations with $V=0$ and since $V_{\rm xc}[n,0]=0$ 
(for otherwise there would be a current flowing in the system at
zero bias, in contradiction to the theorem) 
we have $\d V_{\rm xc}[n,I]/\d 
n(\br) |_{V=0}=\d V_{\rm xc}[n,0]/\d n(\br) = 0$.
Therefore 
\be
G=\frac{G_{s}}{1-G_{s}\left.\frac{\de V_{\rm xc}}{\de 
I}\right|_{I=0}}.
\label{idftG0}
\ee
The i-DFT correction to $G_{s}$ is physically more transparent than 
the TDDFT correction
involving the zero-momentum and zero-frequency limit of the xc kernel 
\cite{sk.2011,ks.2013,ewk.2006,skgr.2007}. 

Although i-DFT has been formulated in $\br$-space the same 
theorem and KS procedure apply to tight-binding models 
by replacing $\br$ with 
a site or orbital index. To highlight the distinctive features of the i-DFT 
potentials in the CB regime, hence above the Kondo temperature, we apply the 
formalism to a junction described by the Constant Interaction Model 
(CIM) with Hamiltonian 
$\hat{H}=\sum_{i\s}\ve_{i}\hat{n}_{i\s}+\frac{1}{2}U\sum_{i\s\neq 
j\s'}\hat{n}_{i\s}\hat{n}_{j\s'}$, 
where $\hat{n}_{i\s}$ is the occupation operator of the $i$-th level 
with spin $\s$.\cite{rate-paper1,obh.2000}
As the electron-electron interaction is confined to the molecular 
junction the Hartree potential vanishes in the leads.
If we choose the region $\mathcal{R}$ as the set of 
interacting levels then the i-DFT theorem states that there is a 
one-to-one map between the pair $(\{\ve_{i}\},V)$ and the pair 
$(\{n_{i}\},I)$.

{\bf Anderson model}
 The CIM with one level  
coupled to two featureless leads, 
$\G_{\a}(\w)=\g/2$, is equivalent to the Anderson model.
Due to spin-degeneracy the i-DFT potentials depend on the 
particle number $N=n_{1\ua}+n_{1\da}$ and current $I$. For
$\ve_{1}=v$, these
can be calculated from 
\begin{subequations}
\bea
N&=&\int [f(\w-V/2)+f(\w+V/2)]A(\w-v)
\\
I&=&\frac{\g}{2}\int 
[f(\w-V/2)-f(\w+V/2)]A(\w-v),
\eea
\label{AMmap}
\end{subequations}
with $A(\w)$  the interacting 
spectral function. Above the Kondo temperature $T_{K}$ a good approximation 
for the spectral function is 
$A(\w)=\frac{N}{2}\ell(\w-U)+(1-\frac{N}{2})\ell(\w)$ with
$\ell(\w)=\g/(\w^{2}+\g^{2}/4)$ \cite{jauho-book}. 
We verified (not shown) that $N$ and $I$ are in excellent agreement with 
the results of the Rate Equations~\cite{rate-paper1,rate-paper2} (RE), the 
orthodox theory of CB for weakly coupled systems. 
For the Anderson model the map $(v,V)\to (N,I)$ is invertible for all 
$V$ (infinite bias window) since the finite-bias Jacobian is
\be
J_{V}=-2\g\,
\frac{a_{+}a_{-}}{2-b_{+}-b_{-}}
\ee
where $a_{\pm}=\int f'(\w\pm V/2)A(\w-v)$ and 
$b_{\pm}=\int f(\w)
[\ell(\w-v-U\mp V/2)-\ell(\w-v\mp V/2)]$. As $a_{\pm}<0$ 
and $-1<b_{\pm}<0$ we find $J_{V}<0$ \cite{observation}. 

\begin{figure}[tbp]
\includegraphics[width=0.48\textwidth]{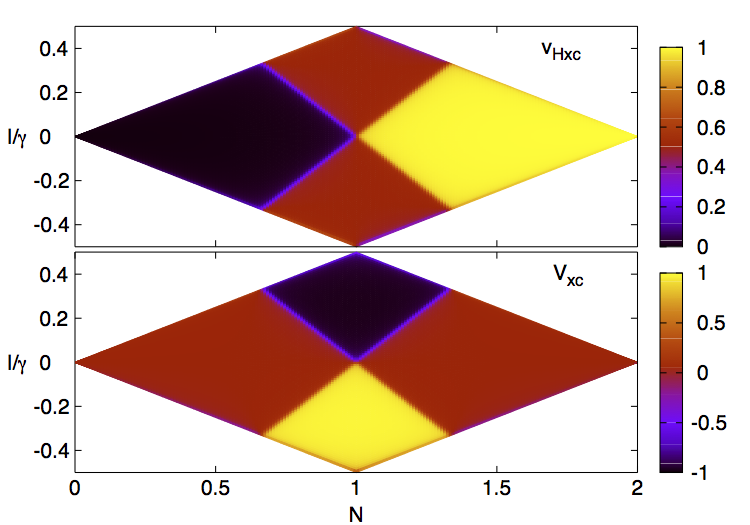}
\caption{Hartree xc gate (top) and xc bias (bottom) of the Anderson model. Energies 
are in units of $U$.}
\label{xcpot-fig}
\end{figure}

To find the xc potentials in Eqs. (\ref{xcpot}) we invert the 
map $(v,V)\to (N,I)$ of Eqs.~(\ref{AMmap}) for $U\neq 0$ and for $U=0$. In terms of the variables 
$w_{\pm}=v\pm V/2$ the problem is separable since the map 
reads $N\mp 2I/\g=2\int f(\w)A(\w-w_{\pm})\equiv \mathfrak{n}(w_{\pm})$, which we solve by 
the bisection method. The values of $\mathfrak{n}\in[0,2]$ and 
therefore  the domain spanned by the particle number and 
current is $|I|\leq\frac{\g}{2}N$ for $N\in [0,1]$ and 
$|I|\leq\frac{\g}{2}(2-N)$ for  $N\in[1,2]$. The reverse engineered xc 
potentials $v_{\rm Hxc}$ and $V_{\rm xc}$ are shown in Fig. 
\ref{xcpot-fig} and 
three observations arise. (i) The Hxc gate (xc bias)
exhibits smeared steps of height $U/2$ ($U$) along the lines $N=1\mp I/\g$. 
Interestingly, the DFT xc discontinuity of $v_{\rm Hxc}[N,0]$ in 
$N=1$ {\em bifurcates} as current starts flowing. (ii) 
The signs of $V_{\rm xc}$ and $I$ are opposite 
(in agreement with the results of Ref.~\citenum{sds.2013}) and the 
derivative $\frac{\de V_{\rm xc}}{\de I}|_{I=0}<0$, thus setting 
$G_{s}$ as the upper limit for the interacting zero-bias conductance [see Eq. 
(\ref{idftG0})]. In Ref.~\citenum{ks.2013} we found that the zero-bias
TDDFT correction at $N=1$ can be expressed in terms of the 
Hxc gate as $\simeq 1/(\frac{2}{\g}G_{s}\frac{\de 
v_{\rm Hxc}}{\de N})$. A comparison with Eq. (\ref{idftG0})  
suggests the existence of a relation (at least in the 
Anderson model) between $V_{\rm xc}$ and 
$v_{\rm Hxc}$ 
since for the two schemes to agree $\frac{\de V_{\rm xc}}{\de I}|_{I=0}=-\frac{2}{\g}\frac{\de 
v_{\rm Hxc}}{\de N}|_{I=0}\gg 1/G_{s}$.
(iii) At finite bias $|V|<U$ and particle number $N=1$ the 
CB prevents current flow in the interacting system. In the same 
situation the KS level pins to the chemical potential and $I$ would be 
large if it were not for the counteraction of the xc bias $V_{\rm 
xc}=-V$. The Landauer+DFT approach~\cite{tgw.2001,bmots.2002,PdC.2004,rgblfs.2006,awe.2007} would therefore fail  
dramatically as
it misses xc bias corrections.

We look for a parametrization of the xc potentials with the 
following properties: ($a$) at zero current $V_{\rm xc}=0$ and $v_{\rm 
Hxc}$ reduces to the Hxc gate of Ref.~\citenum{ks.2013} ($b$) occurrence 
of smeared steps of height $U/2$ for $v_{\rm Hxc}$ (and $U$ for 
$V_{\rm xc}$) at $N=1\mp I/\g$ and ($c$) the derivatives 
$\frac{\de V_{\rm xc}}{\de I}|_{I=0}$ and $\frac{\de v_{\rm xc}}{\de 
N}|_{I=0}$ are related as discussed in observation (iii). The xc 
potentials
\begin{subequations}
\bea
v_{\rm 
Hxc}[N,I]&=&\frac{U}{4}\sum_{s=\pm}\!\left[1+\frac{2}{\p}\,
{\rm atan}\frac{N+\frac{s}{\g}I-1}{W}\right]
\label{xcgatepar}
\\
V_{\rm xc}[N,I]&=&-U\sum_{s=\pm}\frac{s}{\p}\,
{\rm atan}\frac{N+\frac{s}{\g}I-1}{W}\quad\quad
\label{xcbiaspar}
\eea
\label{xcpar}
\end{subequations}

\noindent
have the required properties. 
If we choose the fitting parameter $W\simeq 0.16 \g/U$, 
i-DFT with the xc potentials of Eqs.~(\ref{xcpar}) produces self-consistent 
currents and densities which are almost indistinguishable from the 
interacting ones (not shown), including the density plateaus at 
$\frac{2}{3}$ and $\frac{4}{3}$ as well as the 2:1 ratio of the step 
heights in the current \cite{mgd.2006}. Remarkably, all this is achieved 
without breaking the spin symmetry.

Notice also that in Ref.~\citenum{kskvg.2010} there was no 
xc bias and the current did not reach a steady-state value since the fitting 
parameter $W$ was set to zero (in this case no self-consistent solution of 
the steady-state equation exists).

\begin{figure}[tbp]
\includegraphics[width=0.48\textwidth]{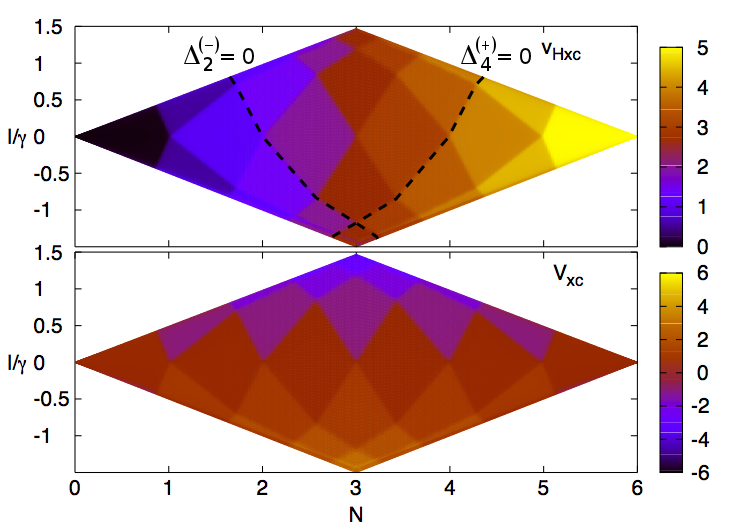}
\caption{Hartree xc gate (top) and xc bias (bottom) of the three-degenerate 
level CIM. The edges of two steps with positive and negative slopes are 
highlighted with dashed lines.
Energies are in units of $U$.}
\label{6levfig}
\end{figure}

{\bf Multiple-level degenerate CIM}
Let us generalize the analysis to a CIM Hamiltonian with 
$\mathcal{M}$ degenerate levels $\ve_{i}=v$ coupled to featureless 
leads, $\G_{\a,ij}(\w)=\d_{ij}\g/2$. Due to degeneracy $v_{\rm Hxc}$ 
is uniform and depends only on 
$N=\sum_{i\s}n_{i\s}$ and 
$I$. Above $T_{K}$ the particle number and current of the interacting and noninteracting 
CIM can be calculated from the RE, 
and then inverted by an adaptation of the iterative bisection 
algorithm of Ref.~\citenum{xctk.2012}. Finally, the i-DFT potentials 
can be obtained 
by subtraction as in Eq.~(\ref{xcpot}).
The map $(v,V)\to (N,I)$ is invertible for 
all $V$ and the codomain is $|I|\leq \frac{\g}{2}N$ for 
$N\in[0,\mathcal{M}]$ and 
$|I|\leq \frac{\g}{2}(2\mathcal{M}-N)$ for 
$N\in[\mathcal{M},2\mathcal{M}]$. The xc potentials are shown in 
Fig.~\ref{6levfig} for $\mathcal{M}=3$. Like in the Anderson model 
the Hxc gate (xc bias) 
exhibits smeared steps of height $U/2$ 
($U$) but the pattern of their edges is  more 
complex. The equilibrium xc discontinuities of $v_{\rm Hxc}[N,0]$ 
at integer $N$ bifurcate with $N$-dependent slopes, and at 
every high-current intersection the slopes of the step edges change. 

In the attempt of 
disentangling the intricate pattern of discontinuity-lines we realized the existence of a duality between 
intersections in the xc potentials and plateaus in the particle number and current. 
From the RE of the degenerate CIM   
a plateau is uniquely identified by a couple of integers $(m,n)$
with $m,n=0,\ldots,2\mathcal{M}$; hence the number of 
plateaus is correctly given by $(2\mathcal{M}+1)^{2}$. In 
the $(m,n)$-plateau with $n\geq m$ the probabilities
$P(q)$, $q=m,\ldots,n$ of 
finding $q$ particles are all identical and given by 
$P^{-1}(q)\equiv P^{-1}_{n\geq m}=\sum_{j=m}^{n}\left(\!\!\begin{array}{c} 2\mathcal{M} \\ j 
\end{array}\!\!\right)$, whereas all other probabilities vanish. The 
corresponding particle number and current are therefore
\begin{subequations}
\bea
N&=&N_{n\geq m}\equiv P_{n\geq m}\sum_{j=m}^{n}j\left(\!\!\begin{array}{c} 2\mathcal{M} \\ j 
\end{array}\!\!\right),
\\
I&=&I_{n\geq m}\equiv\frac{\g}{2}P_{n\geq m}\sum_{j=m}^{n-1}(2\mathcal{M}-j)
\left(\!\!\begin{array}{c} 2\mathcal{M} \\ j 
\end{array}\!\!\right).\quad
\eea
\label{nire}
\end{subequations}
Similarly, for $n\leq m$ one finds $N=N_{n\leq m}=N_{m\geq n}$ and 
$I=I_{n\leq m}=-I_{m\geq n}$. The intersections in the xc 
potentials of Fig.~\ref{6levfig} occur precisely at the points $(N_{n\geq m},I_{n\geq m})$ and 
$(N_{n\leq m},I_{n\leq m})$. Knowledge of these points allows us to 
generalize the parametrization in Eqs.~(\ref{xcpar}) 
\begin{subequations}
\bea
v^{(\mathcal{M})}_{\rm 
Hxc}[N,I]&=&\frac{U}{4}\sum_{K=1}^{2\mathcal{M}-1}\sum_{s=\pm}\!\left[1+\frac{2}{\p}\,
{\rm atan}\frac{\D_{K}^{(s)}(N,I)}{W}\right]\quad\quad
\label{xcgatepar6lev}
\\
V^{(\mathcal{M})}_{\rm 
xc}[N,I]&=&-U\sum_{K=1}^{2\mathcal{M}-1}\sum_{s=\pm}\frac{s}{\p}\,
{\rm atan}\frac{\D_{K}^{(s)}(N,I)}{W}\quad\quad
\label{xcbiaspar6lev}
\eea
\label{xcpar6lev}
\end{subequations}

\noindent
where $\D_{K}^{(s)}(N,I)$ is the piece-wise linear function of $N$ and $I$ 
which vanishes along the step edge passing through $(K,0)$ and having 
positive ($s=+$) or negative 
($s=-$) slopes (for examples see dashed lines in top panel of 
Fig.~\ref{6levfig}). 
 Here $W$ is the same fitting parameter as in Eq.(\ref{xcpar}).
We verified (not shown) that the self-consistent 
solution of the KS equations with 
the xc-potentials of Eqs.~(\ref{xcpar6lev}) are in 
excellent agreement with the RE results.

Using the analytic parametrization of the xc bias in 
Eq.~(\ref{xcbiaspar6lev}) we can calculate the  
zero-bias conductance 
from Eq.~(\ref{idftG0}). The result is
\be
\frac{G}{G_{s}}=\frac{1}{1+\frac{2UG_{s}}{\g \pi W}
\sum_{K=1}^{2\mathcal{M}-1} \frac{\frac{1}{2\mathcal{M}-K+1} + 
\frac{1}{K+1}}{1+ (\frac{N-K}{W})^2}}.
\label{multcond}
\ee
The correction to $G_{s}$ is large for integer $N$'s and Eq.~(\ref{multcond}) 
can be approximated as $\frac{G}{G_{s}}\simeq 1/[1+\frac{2UG_{s}}{\g \pi W}
(\frac{1}{2\mathcal{M}-N+1} + 
\frac{1}{N+1})]$. Thus, the height of the conductance 
peaks depends on the number of electrons in the junction; the closer we get to 
half-filling the larger the height is.

{\bf Multiple-level CIM and finite bias conductance}
The potentials in Eq.~(\ref{xcpar6lev}) 
are not suited to study 
junctions with nondegenerate levels as the dependence on the 
local occupations cannot be reduced to a dependence on $N$ only.
New interesting aspects arise 
which are best illustrated in a HOMO-LUMO CIM with energies 
$\e_{i}=\e_{H/L}+v$ of degeneracy $\mathcal{M}_{H/L}$. 
Let $N_{H/L}$ 
be the particle number on the HOMO/LUMO level, and $v_{\rm 
Hxc}[N_{H},N_{L},I](H/L)$ and $V_{\rm xc}[N_{H},N_{L},I]$ be the 
HOMO/LUMO Hxc gate and xc bias respectively.  
For $N_{L}=0$ (empty LUMO) we have $v_{\rm Hxc}(H)=v_{\rm Hxc}(L)$ 
 (uniform Hxc gate) and the i-DFT 
potentials are given by Eqs.~(\ref{xcpar6lev}) with 
$\mathcal{M}=\mathcal{M}_{H}$. Similarly for 
$N_{H}=2\mathcal{M}_{H}$ (full HOMO) the Hxc gate is uniform and the i-DFT 
potentials are again given by Eqs.~(\ref{xcpar6lev}) but with
$\mathcal{M}=\mathcal{M}_{L}$. At zero current $N_{H}\simeq N$ and 
$N_{L}\simeq 0$ or $N_{H}\simeq 2\mathcal{M}_{H}$ and $N_{L}\simeq N-2\mathcal{M}_{H}$ are the only 
physically realizable occupations. 
Hence, at least for small currents, $v_{\rm Hxc}$ is uniform~\cite{sk.2013} and can 
be parametrized by combining the Hxc gate of two CIMs with degeneracy
$\mathcal{M}=\mathcal{M}_{H/L}$. This argument remains valid for an 
arbitrary number of levels; below we therefore show how to construct 
the i-DFT potentials for the general case.

Let $n=\{n_{1}\ldots 
n_{\mathcal{M}}\}$ be the occupations of levels 
$\{1\ldots\mathcal{M}\}$,
$\mathcal{M}_p[n]$ the degeneracy
of the $p$-th largest occupation and $\mathcal{D}[n]$ the number of 
distinct densities. For instance if $\mathcal{M}=5$ and
$n=\{\frac{1}{3},\frac{1}{2},\frac{1}{2},\frac{1}{3},\frac{1}{3}\}$ then 
$\mathcal{M}_{1}=2$, $\mathcal{M}_{2}=3$ and $\mathcal{D}=2$. 
We further define 
$\mathcal{N}_{p}[n]=2\sum_{q=1}^{p-1}\mathcal{M}_{q}[n]$ as the 
maximum number of particles in the first $(p-1)$ degenerate levels 
($\mathcal{N}_{1}=0$).
The degeneracies $\mathcal{M}_{p}$ are used to 
construct the following i-DFT potentials
\begin{subequations}
\bea
v_{\rm 
Hxc}[n,I]=\sum_{p=1}^{\mathcal{D}[n]}v_{\rm Hxc}^{(\mathcal{M}_{p}[n])}
\big[N-\mathcal{N}_{p}[n],I\big]\quad\quad\quad\quad\quad\quad
\nonumber \\
+\frac{U}{4}\sum_{p=1}^{\mathcal{D}[n]-1}\sum_{s=\pm}\!\left[1+\frac{2}{\p}\,
{\rm 
atan}\frac{N+\frac{2s}{\g}I-\mathcal{N}_{p+1}[n]}{W}\right],\quad\quad
\label{xcgatepar2}
\\
V_{\rm xc}[n,I]=\sum_{p=1}^{\mathcal{D}[n]}V_{\rm xc}^{(\mathcal{M}_{p}[n])}
\big[N-\mathcal{N}_{p}[n],I\big]\quad\quad\quad\quad\quad\quad
\nonumber \\
-U\sum_{p=1}^{\mathcal{D}[n]-1}\sum_{s=\pm}\frac{s}{\p}\,
{\rm atan}\frac{N+\frac{2s}{\g}I-\mathcal{N}_{p+1}[n]}{W}.\quad\quad
\label{xcbiaspar2}
\eea
\label{xcpar2}
\end{subequations}
where, again, $W$ is defined as before.

The dependence on the local occupations enters exclusively through the 
$\mathcal{M}_{p}$. At the joining points ($N=\mathcal{N}_{p+1}[n]$ and 
$I=0$) between two consecutive 
$v_{\rm Hxc}^{(\mathcal{M}_{p}[n])}$ we add a discontinuity with 
slopes $\pm 2/\g$. In fact, the slope of the lines delimiting the 
domain of the i-DFT potentials of a $\mathcal{M}$-fold degenerate CIM are independent of 
$\mathcal{M}$, see Figs.~\ref{xcpot-fig} and \ref{6levfig}. The i-DFT 
potentials in Eqs.~(\ref{xcpar2}) reduce to those in 
Eqs.~(\ref{xcpar6lev}) for equal occupations $n_{i}=N/\mathcal{M}$ 
and to the equilibrium DFT potentials of Ref.~\citenum{ks.2013} for 
$I=0$. Furthermore, they can easily be generalized to CIM with local interactions 
$\frac{1}{2}\sum_{i\s\neq j\s'}U_{ij}\hat{n}_{i\s}\hat{n}_{j\s'}$ and 
level-dependent broadenings $\G_{\a,ij}=\d_{ij}\g_{i}/2$.

\begin{figure}[tbp]
\includegraphics[width=0.48\textwidth]{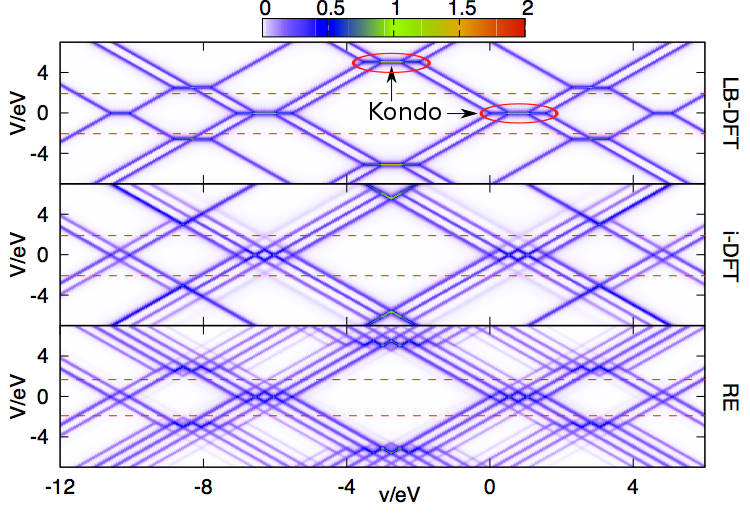}
\caption{Differential conductance (in units of the quantum of conductance) 
calculated from  Landauer+DFT (top), i-DFT 
(middle) and  RE (bottom), for a six-level CIM modelling 
benzene. Dashed red lines delimit the low-bias region where i-DFT and 
RE agree.}
\label{benzene}
\end{figure}

To demonstrate the improvement of i-DFT over Landauer+DFT we calculate
the finite-bias differential conductance $dI/dV$ of a benzene 
junction and benchmark the results against the RE. The benzene is 
described by a six-level CIM with $U=0.5$ 
eV and $\ve_{i}=\ve_{i}^{0}+v$ where 
$\ve^{0}_{1}=-\ve^{0}_{6}=5.08$ eV, 
$\ve^{0}_{2}=\ve^{0}_{3}=-\ve^{0}_{4}=-\ve^{0}_{5}=-2.54$ eV.~\cite{tb-benzene} 
In Fig.~\ref{benzene} we report the $dI/dV$ 
in the three approaches.
Above $T_{K}$ no Kondo plateau is expected. The Landauer+DFT 
scheme (top panel) does instead produce  Kondo plateaus at 
zero~\cite{sk.2011,blbs.2012,tse.2012} and {\em nonzero} biases.
Furthermore, it misses several $dI/dV$ lines as compared to RE 
(bottom panel), thus providing a completely erroneous description of CB.
The i-DFT scheme (middle panel) correctly suppresses the spurious 
Landauer+DFT Kondo plateaus, and reproduces the RE trend of 
the CB peak-heights, see Eq.~(\ref{multcond}) and discussion below. 
At small bias, i-DFT captures {\em all} the $dI/dV$ lines present in 
the RE approach while at higher bias some lines are missing. This latter fact 
is not surprising as our model xc potentials is designed to be 
accurate only at small currents. 

In general, the RE approach requires the solution of a linear 
system of size $4^\mathcal{M}$. In contrast, 
i-DFT requires the solution of $\mathcal{M}$ coupled nonlinear equations. 
If one aims only at describing the low-bias features correctly, 
in i-DFT the solution of two coupled equations is sufficient, 
independent of the size of the system.

{\bf Conclusions}
We propose the i-DFT framework to calculate the steady 
density and current  of interacting junctions at finite bias.
i-DFT is based on the invertibility of the map between $(n,I)$ and $(v,V)$ 
in a finite bias window. 
Unlike the Landauer+DFT approach, i-DFT naturally leads to 
xc bias corrections, and in contrast to TD(C)DFT, the i-DFT xc potentials 
are history-independent. We unveil the complex structure of the i-DFT xc 
potentials in the CB regime by reverse engineering the occupations and 
current of a CIM as obtained from the RE. We found that the bifurcations of the 
xc discontinuity as current starts flowing are pivotal for the 
correct description of CB. Similarly to the equilibrium xc 
discontinuity of standard DFT the bifurcations are an 
intrinsic property of the i-DFT potentials, and are expected to occur 
in the $\br$-space formulation as well. 

We also find an 
efficient parametrization of the i-DFT potentials for small currents 
and use it to calculate the finite bias conductance of a model 
benzene junction. Compared to Landauer+DFT, 
i-DFT shows a clear improvement, being able to reproduce all the 
(small bias) $dI/dV$ lines of the RE approach.


S.K. acknowledges funding by a grant of the "Ministerio de Economia y 
Competividad (MINECO)" (FIS2013-43130-P) and 
by the ``Grupos Consolidados UPV/EHU del Gobierno Vasco'' (IT578-13). 
G.S. acknowledges funding by MIUR FIRB Grant
No. RBFR12SW0J.
We acknowledge support through travel grants [Psi-K2 6084 and 6950 (S.K.) 
and Psi-K2 6468 (G.S.)]
of the European Science Foundation (ESF).


\begin{thebibliography}{46}
\bibitem [{asy()}]{asymbias}%
  \BibitemOpen
  \href@noop {} {}\bibinfo {note} {Since an asymmetric bias $V_{L}\neq - V_{R}$
  between left ($L$) and right ($R$) electrodes is equivalent to a symmetric
  bias $V_{L}=-V_{R}=V/2$ and a uniform shift of the nuclear potential
  $v_{n}(\br)$, we only consider symmetric situations.}\BibitemShut {Stop}%
\end{thebibliography}
\end{document}